\documentclass[12pt]{article}
\usepackage{amsmath,amsfonts,amssymb,amsthm}
\usepackage{graphicx}
\newcommand{\au}{\alpha}
\newcommand{\ad}{\dot{\alpha}}
\newcommand{\bu}{\beta}
\newcommand{\bd}{\dot{\beta}}

\newcommand{\frot}{\frac{1}{2}}
\newcommand{\frof}{\frac{1}{4}}
\newcommand{\froe}{\frac{1}{8}}
\newcommand{\fros}{\frac{1}{16}}
\newcommand{\frit}{\frac{i}{2}}
\newcommand{\frif}{\frac{i}{4}}
\newcommand{\fris}{ \frac{ i }{ \sqrt{2} }}
\newcommand{\frst}{\frac{ \sqrt{2} }{2} }
\newcommand{\frse}{\frac{ \sqrt{2} }{8} }

\newcommand{\OH}{\theta}
\newcommand{\OHh}{\hat{\theta}}
\newcommand{\OHha}{\hat{\theta}^{\alpha}}
\newcommand{\OHhb}{\hat{\theta}^{\beta}}
\newcommand{\OHB}{\bar{\theta}}
\newcommand{\OHBh}{\hat{\bar{\theta}}}
\newcommand{\OHBha}{\hat{\bar{\theta}}^{\dot{\alpha}}}
\newcommand{\OHBhb}{\hat{\bar{\theta}}^{\dot{\beta}}}

\newcommand{\Oa}{\theta^{\alpha}}
\newcommand{\Ob}{\theta^{\beta}}
\newcommand{\Og}{\theta^{\gamma}}

\newcommand{\Os}{\theta^{\sigma}}
\newcommand{\Oad}{\theta_{\alpha}}
\newcommand{\Obd}{\theta_{\beta}}
\newcommand{\Ogd}{\theta_{\gamma}}

\newcommand{\OBa}{\bar{\theta}^{\dot{\alpha}}}
\newcommand{\OBb}{\bar{\theta}^{\dot{\beta}}}
\newcommand{\OBg}{\bar{\theta}^{\dot{\gamma}}}

\newcommand{\OBs}{\bar{\theta}^{\dot{\sigma}}}
\newcommand{\OBad}{\bar{\theta}_{\dot{\alpha}}}
\newcommand{\OBbd}{\bar{\theta}_{\dot{\beta}}}
\newcommand{\OBgd}{\bar{\theta}_{\dot{\gamma}}}

\newcommand{\OO}{\theta\theta}
\newcommand{\OOB}{\bar{\theta} \bar{\theta}}
\newcommand{\OOBB}{\theta \theta \bar{\theta} \bar{\theta}}
\newcommand{\OmO}{\theta\sigma^{m}\bar{\theta}}
\newcommand{\OmOE}{\theta\sigma^{\mu}\bar{\theta}}

\newcommand{\pOa}{\frac{\partial}{\partial \theta^{\alpha}}}
\newcommand{\pOad}{\frac{\partial}{\partial \bar{\theta}^{\dot{\alpha}}}}

\newcommand{\pa}{\partial_{\alpha} }
\newcommand{\pb}{\partial_{\beta} }

\newcommand{\pad}{\partial_{\dot{\alpha}} }
\newcommand{\pbd}{\partial_{\dot{\beta}} }

\newcommand{\pp}{\partial^2}
\newcommand{\pM}{\partial_{m} }
\newcommand{\pME}{\partial_{\mu} }
\newcommand{\pN}{\partial_{n} }
\newcommand{\pL}{\partial_{l} }

\newcommand{\yb}{\bar{y}}
\newcommand{\yh}{\hat{y}}
\newcommand{\ybh}{\hat{\bar{y}}}

\newcommand{\Am}{A_{\mu}}
\newcommand{\An}{A_{\nu}}
\newcommand{\Amu}{A^{\mu}}
\newcommand{\vm}{v_{m}}
\newcommand{\vn}{v_{n}}

\newcommand{\vk}{v_{k}}
\newcommand{\vmu}{v^{m}}

\newcommand{\pmvn}{\partial_{m} v_{n}}
\newcommand{\pnvm}{\partial_{n} v_{m}}
\newcommand{\pkvl}{\partial_{k} v_{l}}

\newcommand{\lam}{\lambda}
\newcommand{\lamB}{\bar{\lambda}}
\newcommand{\lama}{\lambda_{\alpha}}
\newcommand{\lamb}{\lambda_{\beta}}
\newcommand{\lamau}{\lambda^{\alpha}}
\newcommand{\lambu}{\lambda^{\beta}}
\newcommand{\lamgu}{\lambda^{\gamma}}
\newcommand{\lamBa}{{\bar{\lambda}}_{\dot{\alpha}}}
\newcommand{\lamBb}{{\bar{\lambda}}_{\dot{\beta}}}
\newcommand{\lamBg}{{\bar{\lambda}}_{\dot{\gamma}}}

\newcommand{\lamBau}{{\bar{\lambda}}^{\dot{\alpha}}}
\newcommand{\lamBbu}{{\bar{\lambda}}^{\dot{\beta}}}
\newcommand{\lamBgu}{{\bar{\lambda}}^{\dot{\gamma}}}

\newcommand{\DC}{(D-i\partial_{m}v^{m})}
\newcommand{\DCA}{(D-i\partial_{\mu}A^{\mu})}

\newcommand{\Eab}{\epsilon_{\alpha\beta}}
\newcommand{\Eba}{\epsilon_{\beta\alpha}}
\newcommand{\Eag}{\epsilon_{\alpha\gamma}}
\newcommand{\Edb}{\epsilon_{\delta\beta}}
\newcommand{\Ebd}{\epsilon_{\beta\delta}}

\newcommand{\Eabu}{\epsilon^{\alpha\beta}}
\newcommand{\Ebau}{\epsilon^{\beta\alpha}}

\newcommand{\Esbu}{\epsilon^{\sigma\beta}}
\newcommand{\Esgu}{\epsilon^{\sigma\gamma}}

\newcommand{\EBagd}{\epsilon_{\dot{\alpha}\dot{\gamma}}}

\newcommand{\EBbdd}{\epsilon_{\dot{\beta}\dot{\delta}}}

\newcommand{\EBabd}{\epsilon_{\dot{\alpha}\dot{\beta}}}

\newcommand{\EBab}{\epsilon^{\dot{\alpha}\dot{\beta}}}
\newcommand{\EBba}{\epsilon^{\dot{\beta}\dot{\alpha}}}

\newcommand{\Cab}{C^{\alpha\beta}}

\newcommand{\Cgd}{C^{{\gamma\delta}}}

\newcommand{\Cbg}{C^{{\beta\gamma}}}
\newcommand{\Cgb}{C^{{\gamma\beta}}}

\newcommand{\Cmn}{C^{mn}}
\newcommand{\CmnE}{C^{{\mu}{\nu}}}
\newcommand{\CBmn}{{\bar{C}}^{mn}}

\newcommand{\CBab}{\bar{C}^{\dot{\alpha}\dot{\beta}}}
\newcommand{\CBgd}{\bar{C}^{\dot{\gamma}\dot{\delta}}}

\newcommand{\CBgb}{\bar{C}^{\dot{\gamma}\dot{\beta}}}
\newcommand{\CBbg}{\bar{C}^{\dot{\beta}\dot{\gamma}}}

\newcommand{\SBmab}{{\bar{\sigma}}^{m \dot{\alpha} \beta}}
\newcommand{\SBm }{{\bar{\sigma}}^{m}}
\newcommand{\Smnab}{ {\sigma}^{mn{\beta}}_{\alpha} }
\newcommand{\SBmnab}{ {\sigma}^{mn{\dot{\beta}}}_{\dot{\alpha}} }

\newcommand{\Smaa}{\sigma^{m}_{\alpha \dot{\alpha}}}
\newcommand{\SmaaE}{\sigma^{\mu}_{\alpha \dot{\alpha}}}
\newcommand{\Snaa}{\sigma^{n}_{\alpha \dot{\alpha}}}
\newcommand{\Smbb}{\sigma^{m}_{\beta \dot{\beta}}}
\newcommand{\Snbb}{\sigma^{n}_{\beta \dot{\beta}}}
\newcommand{\Smab}{\sigma^{m}_{\alpha \dot{\beta}}}
\newcommand{\Smag}{\sigma^{m}_{\alpha \dot{\gamma}}}
\newcommand{\Smga}{\sigma^{m}_{\gamma \dot{\alpha}}}

\newcommand{\Smba}{\sigma^{m}_{\beta \dot{\alpha}}}
\newcommand{\Snab}{\sigma^{n}_{\alpha \dot{\beta}}}
\newcommand{\Snba}{\sigma^{n}_{\beta \dot{\alpha}}}
\newcommand{\Smsa}{\sigma^{m}_{\sigma \dot{\alpha}}}
\newcommand{\Smsb}{\sigma^{m}_{\sigma \dot{\beta}}}

\newcommand{\Slgb}{\sigma^{l}_{\gamma \dot{\beta}}}
\newcommand{\Smgb}{\sigma^{m}_{\gamma \dot{\beta}}}

\newcommand{\Snbg}{\sigma^{n}_{\beta \dot{\gamma}}}
\newcommand{\Snbs}{\sigma^{n}_{\beta \dot{\sigma}}}
\newcommand{\Slbb}{\sigma^{l}_{\beta \dot{\beta}}}
\newcommand{\Skbg}{\sigma^{k}_{\beta \dot{\gamma}}} 
\newcommand{\Skgg}{\sigma^{k}_{\gamma \dot{\gamma}}}

\newcommand{\SmggE}{\sigma^{\mu}_{\gamma \dot{\gamma}}}
\newcommand{\Skgs}{\sigma^{k}_{\gamma \dot{\sigma}}}
\newcommand{\Sldb}{\sigma^{l}_{\delta \dot{\beta}}}
\newcommand{\Slag}{\sigma^{l}_{\alpha \dot{\gamma}}}

\newcommand{\Qa}{Q_{\alpha}}
\newcommand{\Qb}{Q_{\beta}}
\newcommand{\Qg}{Q_{\gamma}}
\newcommand{\Qd}{Q_{\delta}}
\newcommand{\QBa}{\bar{Q}_{\dot{\alpha}}}
\newcommand{\QBb}{\bar{Q}_{\dot{\beta}}}
\newcommand{\QBg}{\bar{Q}_{\dot{\gamma}}}
\newcommand{\QBd}{\bar{Q}_{\dot{\delta}}}
\newcommand{\Qla}{{\stackrel{\leftarrow}{Q}}_{\alpha}}

\newcommand{\Qlg}{{\stackrel{\leftarrow}{Q}}_{\gamma}}

\newcommand{\QlBa}{{\stackrel{\leftarrow}{\bar{Q}}}_{\dot{\alpha}}}

\newcommand{\QlBg}{{\stackrel{\leftarrow}{\bar{Q}}}_{\dot{\gamma}}}

\newcommand{\Qrb}{{\stackrel{\rightarrow}{Q}}_{\beta}}

\newcommand{\Qrd}{{\stackrel{\rightarrow}{Q}}_{\delta}}

\newcommand{\QrBb}{{\stackrel{\rightarrow}{\bar{Q}}}_{\dot{\beta}}}

\newcommand{\QrBd}{{\stackrel{\rightarrow}{\bar{Q}}}_{\dot{\delta}}}

\newcommand{\Da}{D_{\alpha}}
\newcommand{\Db}{D_{\beta}}
\newcommand{\Dau}{D^{\alpha}}

\newcommand{\DBa}{\bar{D}_{\dot{\alpha}}}
\newcommand{\DBb}{\bar{D}_{\dot{\alpha}}}
\newcommand{\DBau}{ { \bar{D} }^{\dot{\alpha} } }

\newcommand{\covDm}{{\mathcal{D}}_m}
\newcommand{\covDmu}{{\mathcal{D}}^m}

\newcommand{\Phid}{\bar{\Phi}}
\newcommand{\PsiB}{\bar{\psi}}
\newcommand{\PsiBb}{{\bar{\psi}}_{\dot{\beta}}}

\newcommand{\Psib}{{\psi}_{\beta}}

\newcommand{\Psiau}{{\psi}^{\alpha}}
\newcommand{\Psibau}{{\bar{\psi}}^{\dot{\alpha}}}
\newcommand{\Ab}{\bar{A}}
\newcommand{\Fb}{\bar{F}}

\newcommand{\Wa}{W_{\alpha}}
\newcommand{\WBa}{{\overline{W}}_{\dot{\alpha}}}

\newcommand{\WB}{\overline{W}}

\newcommand{\DT}{\delta}

\newcommand{\PhiB}{\bar{\Phi}}
\newcommand{\Lam}{\Lambda}
\newcommand{\LamB}{\bar{\Lambda}}
\newcommand{\pMp}{\partial_{m} \phi }
\newcommand{\pNp}{\partial_{n} \phi }
\newcommand{\ppp}{\partial^2 \phi }
\newcommand{\Lamp}{\Lam_p}
\newcommand{\LamBp}{\bar{\Lam}_p}
\newcommand{\bob}{{\epsilon}_{\alpha\beta}C^{\beta\gamma}
                  {\sigma}^{m}_{\gamma{\dot{\alpha}}}}
\newcommand{\jane}{{\bar{\epsilon}}_{{\dot{\alpha}}{\dot{\beta}}}
                  {\bar{C}}^{{\dot{\beta}}{\dot{\gamma}}}
                  {\sigma}^{m}_{\alpha{\dot{\gamma}}}}
\newcommand{\eF}{{\epsilon}^F}
\newcommand{\eG}{{\epsilon}^G}

\newcommand{\eFG}{{\epsilon}^{FG}}

\newcommand{\Fmn}{{F}_{mn}}
\newcommand{\Fmnu}{{F}^{mn}}
\newcommand{\Flku}{{F}^{lk}}
\newcommand{\Emnlk}{{\epsilon}_{mnlk}}

\begin{document}

\title{Gauged Wess-Zumino Model in Noncommutative\\ Minkowski Superspace}
\author{James S. Cook \\ \\ 
        \it{Department of Mathematics}\\
        \it{North Carolina State University}\\
        \it{Raleigh NC 27695}}
\date{May 27, 2005}

\renewcommand{\thefootnote}{\fnsymbol{footnote}}
\footnotetext[0]{ a similar paper has been submitted to JHEP } 
\renewcommand{\thefootnote}{\fnsymbol{footnote}}
\footnotetext[0]{testbetter@yahoo.com}

\maketitle 
 
\abstract{
We develop a gauged Wess-Zumino model in noncommutative Minkowski superspace. 
This is the natural extension of the work of Carlson and Nazaryan, which extended $N=1/2$ supersymmetry written over deformed Euclidean superspace to Minkowski superspace. We investigate the interaction of the vector and chiral superfields. Noncommutativity is implemented 
by replacing products with star products.  Although, in general, our star product is nonassociative, we prove that it is associative to the first order in the deformation parameter $C$.  We show that our model reproduces the $N=1/2$ theory in the appropriate limit, namely when the deformation parameters $\CBab = 0$.  Essentially,
we find the $N=1/2$ theory and a conjugate copy.  As in the $N=1/2$ theory, a reparameterization of the gauge
parameter, vector superfield and chiral superfield are necessary to write standard C-independent gauge theory.
However, our choice of parameterization differs from that used in the $N=1/2$ supersymmetry, which leads to some unexpected new terms. 
}

\vfill

\pagebreak


\section{Overview}
There has been a variety of discussion about deformations of superspace in recent years \cite{peter}-\cite{sasaki} (undoubtedly a partial list). Of particular interest to this paper is the deformed Euclidean superspace constructed by Seiberg in 
\cite{sb03}. Generally, the literature following Seiberg has focused on  
superspace with a Euclidean signature. One exception is \cite{naz}, in which Carlson and Nazaryan found how to construct a deformed Minkowski superspace
\footnote{After the completion of this work the author learned that 
the work of M. Chaichian and A. Kobakhidze in \cite{archil} and the work of 
Y. Kobayashi and S. Sasaki in \cite{sasaki} also studied the Wess-Zumino model on deformed Minkowski superspaces in some detail. Both of these works employ a star product which is associative but not hermitian. The star product studied here is hermitian but not associative in general. Also, note that \cite{klemm} and \cite{ferraraII} study some aspects of deformed Minkowski superspace that have relevance to this work.}.
In their paper, they implemented superspace noncommutativity with a star 
product which was hermitian but not associative in general. Their star product reproduces the deformation of $N=\frot$ supersymmetry in a certain limit. Additionally, they studied the Wess-Zumino model (without gauge interactions) and found results similar to Seiberg's. Our goal is to construct the gauged Wess-Zumino model in this noncommutative Min\-kow\-ski superspace.

Following the construction of Nazaryan and Carlson, we deform $N=1$ rigid Minkowski superspace as follows:
\begin{equation} \label{E:deformation}
\begin{array}{l}
      \{ \OHha , \OHhb \} = \Cab \qquad \{ \OHBha , \OHBhb \} = \CBab  
\end{array}
\end{equation}
where $(\Cab)^* = \CBab $. In this deformation, all of the fermionic dimensions of superspace are deformed. Here both $Q$ and $\bar{Q}$ are broken symmetries, so we will say that this space has $N=0$ supersymmetry. Despite this, the deformation still permits most of the usual superfield constructions.

In section \ref{S:hats} we explicitly define the noncommutative Minkowski superspace by listing the deformed coordinate algebra found in \cite{naz}. The deformed coordinates have hats on them to distinguish them from the standard coordinates.
The usual model is then deformed by simply putting a hat on all of the objects in the standard theory. In practice, we will not explicitly calculate anything in terms of these operators. Instead, 
we will find it useful to make the usual exchange of the operator product for the star product of ordinary functions of superspace; 
\begin{equation} \label{E:mapping}
\begin{array}{l}
      \hat{f_1}\hat{f_2}\mapsto f_1 \ast f_2. 
\end{array}
\end{equation}
This correspondence allows us to work out the details of noncommutative theory using ordinary calculus on superspace. In this sense we obtain the 
noncommutative Wess-Zumino model by simply replacing ordinary products with star products.

In sections \ref{S:starproduct} and \ref{S:supersymmetry}, we continue our brief summary of the work of Carlson and Nazaryan in \cite{naz}. In \cite{naz}, deformed Minkowski space was constructed to the second order in the deformation parameter. In this paper, we will primarily examine the first order extensions of their work.  In section \ref{S:vector}, we will examine how to write a nonabelian supersymmetric gauge theory on noncommutative Minkowski superspace. Following the standard superfield construction(see \cite{wb} for example), we will introduce the vector superfield (V) and calculate the star exponential ($e^V$) in section \ref{S:exponential}. We will calculate the explicit modification these definitions imply for the component fields of the vector multiplet.

The gauge transformation itself will be discussed in section \ref{S:gauge}. In section \ref{S:gauge transformations}, we will find a parameterization of the vector superfield such that the standard gauge transformations are realized at the component field level. This procedure is similar to Seiberg's in \cite{sb03}. We employ a modified Wess-Zumino gauge throughout the calculations. This is possible provided that we define the gauge parameter $\Lambda$ with some carefully chosen deformation dependent shifts. We will find that reality uniquely affixes this construction. Next, in section \ref{S:field strength}, we introduce the spinor superfield $\Wa$ by making the natural modification to the standard definition.

Then, in section \ref{S:matter fields}, we examine the gauge transformation on a chiral superfield. Again, we will find it necessary to shift the chiral superfield by a deformation dependent term in order to preserve the usual gauge theory. These shifts, similar to those found in \cite{sb03} and \cite{aio}, are derived in detail.

Finally, in section \ref{S:lagrangian}, we construct the Lagrangian of the gauged Wess-Zumino model.  This construction closely resembles that of Wess and Bagger in \cite{wb} except that products have been replaced by star products.  Also, the component field expansions of the superfields have some C-dependent shifts as derived in the previous sections.  Overall, the gauge symmetry of the Lagrangian is established by arguments analogous to the standard arguments.  We conclude the paper by computing the Lagrangian written explicitly in terms of the component fields.  Our result is similar to \cite{aio}, however, there are some unexpected terms.\footnote{This paper contains some second order results for the star exponential. However, we do not complete the development of the theory to second order in this paper. We do find some partial results at the second order of the deformation parameter and they agree with the $N=\frot$ in the limit 
$\CBab=0$. The JHEP version contains only the first order results given here.} 

\section{Noncommutative Minkowski Superspace} 
\subsection{Deformed Coordinate Algebra}\label{S:hats}
We begin by considering $N=1$ rigid Minkowski superspace where a typical point is 
$(x^{m},\Oa,\OBa)$. In the commutative case, we have:
\begin{equation}
\begin{array}{ll}
[x^{m},x^{n}]=0 &\qquad \qquad [x^m , \Oa ]=0\\
\{{\theta}^{\alpha},{\theta}^{\beta}\}=0 &\qquad \qquad [x^{m}, {\bar{\theta}}^{\dot{\alpha}}]=0\\
\{{\bar{\theta}}^{\dot{\alpha}},{\bar{\theta}}^{\dot{\beta}}\}=0 &\qquad \qquad \{{\theta}^{\alpha},{\bar{\theta}}^{\dot{\beta}}\}=0
\end{array}
\end{equation}
The coordinates $x^{m}$ are identified with spacetime coordinates, whereas the $\Oa$ 
and $\OBa$ are grassman variables. We then construct noncommutative Minkowski 
superspace by replacing coordinate functions $(x^{m}$, $\Oa$, $\OBa)$ 
with operators $(\hat{x}^{m}$, $\OHha$, $\OHBha )$. In particular,
we require that the deformed coordinates satisfy
\begin{equation}
\begin{array}{ll}
\{\hat{\theta}^\alpha,\hat{\theta}^\beta\} =C^{\alpha\beta} &\qquad
[\hat{x}^{m},\hat{\theta}^\alpha] = iC^{\alpha\beta}{\sigma}^{m}_{\beta{\dot{\beta}}}\hat{\bar{\theta}}^{\dot{\beta}} 
\\
\{\hat{\bar{\theta}}^{\dot{\alpha}},\hat{\bar{\theta}}^{\dot{\beta}}\}  
= {\bar{C}}^{ \dot{\alpha}\dot{\beta} } &\qquad
[\hat{x}^{m},\hat{\bar{\theta}}^{\dot{\alpha}}]=i{\bar{C}}^{\dot{\alpha}\dot{\beta}}
\hat{\theta}^{\beta}{\sigma}^{m}_{\beta{\dot{\beta}}}
\\
\{\hat{\theta}^\alpha,\hat{\bar{{\theta}}}^{\dot{\beta}}\} = 0 &\qquad
[\hat{x}^m ,\hat{x}^n] = (C^{\alpha\beta}\hat{\bar{\theta}}^{\dot{\alpha}}\hat{\bar{\theta}}^{\dot{\beta}}-{\bar{C}}^{\dot{\alpha}\dot{\beta}}\hat{\theta}^\alpha\hat{\theta}^\beta){\sigma}^{m}_{\alpha{\dot{\alpha}}}{\sigma}^{n}_{\beta{\dot{\beta}}}.
\end{array}
\end{equation}
This algebra was defined by Carlson and Nazaryan so that the deformed 
chiral coordinates 
${\hat{y}}^{m}=\hat{x}^{m}+i{\hat{\theta}}{\sigma}^{m}\hat{\bar{\theta}}$ and 
${\hat{\bar{y}}}^{m}={\hat{x}}^{m}-i{\hat{\theta}}{\sigma}^{m}{\hat{\bar{\theta}}}$ satisfy
\begin{equation}
\begin{array}{ll}
\{\hat{\theta}^\alpha,\hat{\theta}^\beta\} =C^{\alpha\beta} &
[\hat{y}^{m},\hat{\theta}^\alpha] = 0
\\
\{\hat{\bar{\theta}}^{\dot{\alpha}},\hat{\bar{\theta}}^{\dot{\beta}}\}  
= {\bar{C}}^{ \dot{\alpha}\dot{\beta} } &
[\hat{\bar{y}}^{m},\hat{\bar{\theta}}^{\dot{\alpha}}]=0
\\
\{\hat{\theta}^\alpha,\hat{\bar{{\theta}}}^{\dot{\beta}}\} = 0. 
\end{array}
\end{equation}
These relations will allow us to develop chiral and antichiral superfields in much the same 
way as in the commutative theory. In addition, we have: 
\begin{eqnarray}
{[} \ybh^{m}, \OHh^{\au} {]} & 
  = & 2i \Cab \Smbb \OHBh^{\bd} \notag \\ 
{[} \yh^{m}, \OHBh^{\ad} {]} & 
  = & 2i \CBab \OHh^{\bu} \Smbb \notag \\
{[} \yh^{m}, \yh^{n} {]} &
  = & \left( 4\CBab \OHh^{\au} \OHh^{\bu} - 2 \Cab \CBab \right) 
      \Smaa \Snbb \\ 
{[} \ybh^{m}, \ybh^{n} {]} &
  = & \left( 4 \Cab \OHBh^{\ad} \OHBh^{\bd} - 2 \Cab \CBab \right) 
      \Smaa \Snbb \notag \\ 
{[} \yh^{m}, \ybh^{n} {]} &
  = & 2 \Cab \CBab \Smaa \Snbb. \notag  
\end{eqnarray}
This choice of deformed coordinate is motivated by our desire to follow the same construction of chiral superfields as in the commutative theory. 


\subsection{Star Product} \label{S:starproduct}
The star product on Minkowski superspace is defined by 
\begin{equation}
f\ast g = f(1+S)g
\end{equation}
where  $S$ is formed using the supercharges $\Qa$ and $\QBa$, 

\begin{equation}
\begin{array}{ll}
S \ \ = & -\frot \Cab \Qla \Qrb -\frot \CBab \QlBa \QrBb \notag \\
        & + \froe \Cab \Cgd \Qla \Qlg \Qrd \Qrb 
          + \froe \CBab \CBgd \QlBa \QlBg \QrBd \QrBb \\
        & +   \frof \Cab \CBab ( \QlBa \Qla \QrBb \Qrb -
                               \Qla \QlBa \Qrb \QrBb )\notag.
\end{array}
\end{equation}
                            
We follow the conventions of Wess and Bagger in \cite{wb}. In the chiral coordinates $y^{m}=x^m +i\OmO$, the supercharges have the familar forms. Note that the derivatives of $\Oa$ and $\OBa$ are taken at fixed $y^m$. 
\begin{equation}
\begin{array}{l}
\Qa = \pOa |_y \\
\QBa = - \pOad |_y +2i \Oa \Smaa \frac{\partial}{\partial y^{m}}
\end{array}
\end{equation}
Whereas, when the derivatives are taken at fixed antichiral 
coordinates $\bar{y}^m =x^m -i\OmO$, we have
\begin{equation}
\begin{array}{l}
\Qa = \pOa |_y -2i \Smaa \OBa \frac{\partial}{\partial \bar{y}^{m}}\\
\QBa = - \pOad |_{\bar{y}}. 
\end{array}
\end{equation}
We will not make explicit $|_{y}$ or $|_{\bar{y}}$ elsewhere, since they are to be understood implicitly. Many other formulae can be found in \cite{naz}. Some properties of this star product on functions $f$, $g$, and $h$ are
\begin{equation}
\begin{array}{ll}
\overline{f \ast g} = {\overline{g}} \ast {\overline{f}} &\qquad
(f+g)\ast h = f \ast h + g \ast h \\
f \ast g \ne g \ast f &\qquad
f \ast (g \ast h) \ne (f \ast g) \ast h.
\end{array}
\end{equation}
The noncommutativity and nonassociativity will require some attention in general. However, to the first order in the deformation parameter, we note that
\begin{equation} \label{E:assoc}
\begin{array}{l}
f \ast (g \ast h) = (f \ast g) \ast h 
\end{array}
\end{equation}
the star product is associative. A proof is given in the appendix.


\subsection{$N=0$ Supersymmetry} \label{S:supersymmetry}
The formulae below are stated for the operators acting on functions of the
deformed Minkowki superspace. In particular, they should be understood as statements about how the operators act on star products of functions. We define the star brackets as 
\begin{equation}
\begin{array}{ll}
\{A,B\}_{\ast} = A \ast B + B \ast A \ \ &\mathrm{and} \ \ \
{[}A,B{]}_{\ast} = A \ast B - B \ast A.
\end{array}
\end{equation}
Then calculate 
\begin{equation}
\begin{array}{l}
\{\Oa,\Ob\}_{\ast} \   = \Oa \ast \Ob + \Ob \ast \Oa = \Cab \\
\{\OBa,\OBb\}_{\ast} \ = \OBa \ast \OBb + \OBb \ast \OBa = \CBab. \\
\end{array}
\end{equation}
It is important to note that products of both $\Oa$ and $\OBa$ are deformed. This has the consequence of 
breaking all of the supersymmetry. Starting with the canonical forms of the supercharges, we obtain
\begin{equation}
\begin{array}{ll}
\{\Qa,\Qb\}_{\ast}   &= -4\CBab \Smaa \Snbb 
             \frac{\partial^2}{\partial\bar{y}^{m} 
                               \partial\bar{y}^{n}} \\
\{\QBa,\QBb\}_{\ast} &= -4\Cab \Smaa \Snbb 
             \frac{\partial^2}{\partial y^{m} 
                               \partial y^{n}} \\
\{\Qa,\QBa\}_{\ast}  &= 2i \Smaa \frac{\partial}{\partial y^{m}}.
\end{array}
\end{equation}
Comparing this to \cite{sb03}, we note that when $\CBab =0$, then $\Qa$ is an unbroken symmetry, 
hence the label $N=\frac{1}{2}$ supersymmetry. The author proposes that we call the theory 
constucted by Carlson and Nazaryan $N=0$ supersymmetry to be consistent. Now, although the 
supercharges are broken, we still have
\begin{equation} \label{E:QDalgebra}
\begin{array}{c}
\{\Da,\Qb\}_{\ast} = \ \{\DBb,\Qb\}_{\ast} 
                   = \ \{\Da,\QBb\}_{\ast}  
                   = \ \{\DBa,\QBb\}_{\ast} = 0 \\
\ \{\Da,\Db\}_{\ast} = \ \{\DBa,\DBb\}_{\ast} = 0. 
\end{array}
\end{equation}
These relations are crucial. We can still define the chiral $(\Phi)$ and antichiral 
$(\bar{\Phi})$ superfields by the constraints 
$\DBa \ast \Phi = 0$ and $\Da \ast \bar{\Phi} = 0$ on noncommutative Minkowski superspace. Thus, most of the usual techinques in Wess and Bagger \cite{wb} still apply for our discussion. The primary difference is that products will be replaced with star products.


\section{Vector Superfields} \label{S:vector}

Our goal is to construct a nonabelian gauge on deformed Minkowski superspace. Thus, we consider a vector superfield $V$ which carries some matrix representation of the gauge group and is subject to the usual constraint: $\overline{V} = V$. In the standard super Yang-Mills theory, it is convenient to use a reduced set of component fields called the Wess Zumino gauge. We will show in section \ref{S:gauge transformations} that the Wess Zumino gauge can be generalized to the current discussion provided we make some $C$ dependent shifts. For now, we let $V$ take the canonical parametrization of the Wess-Zumino gauge
\begin{equation}
V = -\OmO \vm +i\OO\OHB\lamB -i\OOB\OH\lam + \frot \OO\OOB \DC
\end{equation}
where the above is in chiral coordinates $y^m$.

\subsection{Star exponential of vector superfield} \label{S:exponential}

We define the star exponential of the vector superfield in the natural way:
\begin{equation}
e^V = 1+V+\frac{1}{2}V\ast V+\frac{1}{3!} V\ast V \ast V + ...
\end{equation}
Our notation for the usual exponential will be $exp(V)$ and powers are to be understood as ordinary
powers - for example $V^2 = VV$. In this paper, star products will be explicitly indicated. 


The vector superfield is even, thus no new signs arise from pushing 
the $\Qa$ or $\QBa$ past V in the star product. Thus, to first order in the deformation parameter,
\begin{eqnarray}
V \ast V &= & V(1+S)V \notag \\
         &= & V^2-\frot \Cab (\Qa V)( \Qb V) - \frot \CBab (\QBa V )(\QBb V )\notag \\
         &  &+ \froe \Cab \Cgd (\Qa \Qg V)( \Qd \Qb V) 
	    +   \froe \CBab \CBgd (\QBa \QBg V )(\QBd \QBb V)  \notag \\ 
         &  &+ \frof \Cab \CBab \left( (\QBa \Qa V)( \QBb \Qb V) 
                                       -(\Qa \QBa V)(\Qb \QBb V )\right)\notag                        
\end{eqnarray} 
We will now calculate these terms in chiral coordinates starting with                                                  
\begin{eqnarray} \label{E:QaV}
\Qa V &= & \pa \biggl[ -\OmO \vm +i\OO\OHB\lamB -i\OOB\OH\lam + \frot \OO\OOB \DC \biggr] \notag \\
      &= & -\Smaa \OBa \vm + 2i\Oad \OHB \lamB + \OOB \bigl( -i\lam_{\au} +\Oad \DC \bigr).                     
\end{eqnarray} 
Continuing, we find that
\begin{eqnarray}
\Qb \Qa V &= & \pb  \biggl[  -\Smaa \OBa \vm + 2i\Oad \OHB \lamB + \OOB(-i\lam_{\au} +\Oad \DC \bigr) \biggr] \notag \\
          &= & -2i\Eba \OHB \lamB  - \Eba \OOB \DC.                  
\end{eqnarray} 
Next we calculate $ \QBa V $.
\begin{eqnarray} \label{QBaV}
\QBa V &= & (-\pad +2i\Oa \Snaa \pN )\biggl[ -\OmO \vm +i\OO \OHB \lamB -i\OOB \OH \lam + \frot \OO\OOB \DC \biggr] \notag \\
       &= & -\Oa \Smaa \vm +\bigl(-2i \OBad +2 \OOB \Smaa \Oa \pM \bigr)  \OH \lam \notag \\
       &  &+ \OO \biggl( i \lamBa +\OBad \DC +i\Eabu \Smaa \Snbb \OBb \pmvn \biggr)                  
\end{eqnarray} 
The next calculation is a bit longer.
\begin{eqnarray}
\QBa \QBb V &= & (-\pad +2i\Oa \Smaa \pM )( \QBb V) \\
            &= & -2i \EBabd \OH \lam + \OO 
	          \biggl( \EBabd \DC \notag \\
            &  &+i\Eabu (\Smab \Snba -\Smaa \Snbb )\pmvn 
                +2(\Smaa \OBb -\OBa \Smab )\pM \lamau \biggr) \notag                   
\end{eqnarray} 
Now, for the mixed supercharges, using the results above, we find that
\begin{eqnarray}
\Qa \QBa V &= & \pa (\QBa V) \\
           &= & -\Smaa \vm + 2i(\Oad \lamBa -\OBad \lama) \notag \\
	   &  &\ + \Oad \biggl( 2\OBad \DC +2i \OBb \Esbu \Smsa \Snbb \pmvn 
+2\OOB \Smba \pM \lambu \biggr) \notag          
\end{eqnarray} 
Similarly, we find that
\begin{eqnarray}
\QBa \Qa V &= & (-\pbd +2i\Oa \Smaa \pM )(\Qa V) \\
           &= & \Smaa \vm - 2i(\Oad \lamBa -\OBad \lama) \notag \\
	   &  &\ + 2\Oad \OBad \DC -2i \Ob \OBb \Smba \Snab \pmvn \notag \\
           &  &\ - 2\OO \Smaa \pM (\OHB \lamB) +
               \OOB \biggl( 2\Ob \Smba \pM \lama +i\OO \Smaa \pM \DC 
               \biggr) \notag          
\end{eqnarray} 

The next task is to calculate the products of the terms above. In the product below, we have omitted from the beginning those terms with $\OOB$ because there is a $\OHB$ in each term.
\begin{eqnarray} \label{E:cQVQV}
\frot \Cab \Qa V \Qb V &= &\frot\Cab
     \bigl[ -\Smaa \OBa \vm + 2i\Oad(\OHB \lamB) \bigr]  
     \bigl[ -\Snbb \OBb \vn + 2i\Obd(\OHB \lamB) \bigr]  \\
    &= & \frof \Cab \EBab \Smaa \Snbb \vm \vn \OOB 
         +\frit \Cab \Ob \Smaa [ \vm , \lamBau ] \OOB \notag \\
    &= & \biggl( \frot \Cmn \vm \vn -\frit \Cab \Smaa \OBb [ \vm , \lamBau ] \biggr) \OOB 
\notag
\end{eqnarray} 
where we have used the identity $\Cmn = \frot \Cab \EBab \Smaa \Snbb $, following the conventions of \cite{sb03}. 
Continuing to compute the products, since every term has a $\OH$ this time, we can ignore the $\OO$ terms from the outset.
\begin{eqnarray}
\frot \CBab \QBa V \QBb V &= &\frot\CBab
     \bigl[ -\Oa \Smaa \vm - 2i\OBad \OH \lam \bigr]  
     \bigl[ -\Ob \Snbb \vn - 2i\OBbd \OH \lam  \bigr]  \\
    &= & -\frof \CBab \Ebau \Smab \Snba \vm \vn \OO 
         -\frit \CBab \OBbd \Smaa [ \vm , \lama ] \OO \notag \\
    &= & \biggl( \frot \CBmn \vm \vn +\frit \CBab \Smaa \OBbd [ \vm , \lamau ] \biggr) \OO
\notag
\end{eqnarray} 
where we identified $\CBmn = -\frot \CBab \Eabu \Smaa \Snbb $ following \cite{naz}. Next, consider the second order in deformation parameter terms: 
\begin{eqnarray}
\froe \Cab \Cgd (\Qa \Qg V )(\Qd \Qb V ) &= & 
\froe \Cab \Cgd \Eag \Ebd [ 2i\OHB \lamB + \OOB \DC ]^2 \notag \\
&= & -\froe |C|^2 \lamB \lamB \OOB \notag
\end{eqnarray} 
where we use $|C|^2 = 4\Cab \Cgd \Eag \Edb $ . Similarly, we find that the next term is easily calculated due to a sizeable
cancellation since we may omit a $\OO$ term from the start.
\begin{eqnarray}
\froe \CBab \CBgd (\QBa \QBg V )(\QBd \QBb V ) &= & 
\froe \CBab \CBgd \EBagd \EBbdd [ -2i\OH \lam ]^2 \notag \\
&= &-\froe |\bar{C} |^2 \lam \lam \OO \notag
\end{eqnarray} 
where we use $|\bar{C} |^2 = 4\CBab \CBgd \EBagd \EBbdd $. 
 The remaining term to consider in $V\ast V$ is $\frof \Cab \CBab {[}
(\QBa \Qa V)(\QBb \Qb V) - (\Qa \QBa V)(\Qb \QBb V){]}$. We calculate
\begin{equation} \label{E:mixed}
\begin{array}{l}
\frof \Cab \CBab \biggl( (\QBa \Qa V)(\QBb \Qb V) - 
                      (\Qa \QBa V)(\Qb \QBb V) \biggr) = \\
\qquad = \ \frof \Cab \CBab \biggl( \Smaa 
                     \{ \vm , 4i(\OBb \lamb -\Ob \lamBb )\}  \\
\qquad \qquad - \ 2i\Smaa \{ \vm , \pL \vk (\Og \Slgb \Skbg \OBg +
               \Ob \Esgu \Sldb \Skgg \OBg ) \} \\
\qquad \qquad + \ 2\Smaa \OOB \Slgb \{ \vm , \Og \pL \lamb
                            -\Ob \pL \lamgu  \} \\
\qquad \qquad - \ 2\Smaa \OO \Slbb \{ \vm , \pL ( \OHB \lamB ) \} \\
\qquad \qquad + \ i\Smaa \OOBB \Slbb \{ \vm , \pL \DC \} \\
\qquad \qquad - \ 4 \Oad \Og \Smgb \Snbg \OBg \{ \lamBa , \pmvn \} \\
\qquad \qquad + \ 4 \OBad \Og \Smgb \Snbg \OBg \{ \lama , \pmvn \} \\
\qquad \qquad + \ 4i \Oad \OOB \Og \Smgb \{ \lamBa , \pM \lamb \} \\
\qquad \qquad + \ 4i \Oad \OO \Og \Smbb \{ \lama , \pM \lamBg \} \\
\qquad \qquad - \ 4i \Oad \OBad \Og \Smgb \Snbg \OBg \{ \DC , \pmvn \} \\
\qquad \qquad - \ 4 \Og \OBg \Os \OBs \Skgs \Slag \Smsb \Snbs \pkvl \pmvn 
\biggr). 
\end{array}
\end{equation}
We can see from the expression above the full second order calculations will be lengthy. Additionally, we would have to deal with the nonassociativity of the star product. At present, the 
author has only calculated portions of the theory to the second order, mostly for the purpose of comparing the present work with \cite{sb03}. We leave the complete development of the second order deformed gauge theory to a later paper.


\subsubsection{Expanding $V \ast V \ast V$}
We shall now find the correction to $ V\ast V \ast V $ to the first order in
$\Cab$. First, recall first that in the commutative theory, $V^3$ is zero in the Wess-Zumino gauge. 
Thus any nontrivial term in $ V\ast V \ast V $ must arise from the deformation.
\begin{equation}
\begin{array}{l}
V\ast (V\ast V) \ = \\
\qquad = \ V(V\ast V)-\frot \Cab (\Qa V)\Qb (V\ast V) 
              -\frot \CBab (\QBa V)\QBb (V\ast V)  \\
\end{array}
\end{equation}
We can replace $V\ast V$ with $V^2$ as we are looking for the first order in $\Cab$ terms.
\begin{equation}
\begin{array}{l}
V\ast (V\ast V) \ = \\
\qquad = \ V(V\ast V)-\frot \Cab (\Qa V)\Qb (V^2) 
              -\frot \CBab (\QBa V)\QBb (V^2)  \\
\qquad = \ V(V\ast V) 
\end{array}
\end{equation}
The two terms vanish because $\Qa V$ and $\QBa V$ have a $\OHB$ and $\OH$
in each term respectively while $\Qb (V^2)$ and $\QBb (V^2)$ are proportional to $\OOB$ and $\OO$ respectively. To the first order, we have
\begin{equation}
\begin{array}{l}
V\ast (V\ast V) \ = \\
\qquad = \bigl(-\OmO \vm +i\OO\OHB\lamB -i\OOB\OH\lam + \frot \OO\OOB \DC \bigr) (V\ast V). \\
\end{array}
\end{equation}
Now, if we examine the first order terms in $V\ast V$, we notice that each term either has $\OO$ or $\OOB$; thus, the product 
with $V$ which is proportional to $\OH$ and $\OHB$ vanishes. Therefore, to the first order in the deformation parameter,
\begin{equation}
\begin{array}{l}
V\ast (V\ast V) \ = 0. \\
\end{array}
\end{equation}
It is not hard to see that this extends to higher star products. Thus, ${(V)}_{\ast}^n = 0$ for $n \geq 3 $ to the first order in the deformation parameter. That is, to the first order in C, we have $e^V = 1+V+\frot V\ast V$. This is nice but it will clearly be spoiled if we include the second order terms. For example, if one examines the mixed second 
order term (\ref{E:mixed}), the first few lines have only $\OH$ or $\OHB$. Hence, in the product with V they will not vanish like the first order 
case, thus generating a nontrivial term in $V\ast (V\ast V)$. We will not complete the development of $e^V$ to the second order in 
this paper. Next, we shall show that in the limit of $\CBab=0$, we recover the terms found by Seiberg in \cite{sb03}. 

Collecting the results of this section, we find that the star exponential of V in the canonical Wess-Zumino gauge is
\begin{equation} \label{E:stexp}
\begin{array}{l}
e^V = 1 + V + \frot V\ast V \\
 \ \ \ \ = 1-\OmO \vm +i\OO\OHB\lamB 
              -i\OOB\Oa \lama + \frot \OO\OOB \DC \\ 
 \ \, \, \ \qquad -  \ \ \bigl( \frof \Cmn \vm \vn
            +   \frif \Cab \Obd \Smaa [ \lamBau , \vm ] \bigr) \OOB \\
 \ \, \, \ \qquad - \ \ \bigl( \frof \CBmn \vm \vn
            +   \frif \CBab \OBbd \Smaa [ \vm , \lamau ] \bigr) \OO.  \\
 \ \, \, \ \qquad - \ \fros |C|^2 \lamB \lamB \OOB \\
 \ \, \, \ \qquad - \ \fros |\bar{C} |^2 \lam \lam \OO \\
 \ \, \, \ \qquad + \ other \ 2^{nd} \ order \ terms \ containing  \ \CBab.  \\
\end{array}
\end{equation}


\subsection{ $N=0$ verses $N=\frot$ star exponentials} \label{S:comparison}

To compare with the $N=\frot$ construction, we make the following dictionary: 
\begin{equation} \label{E:dict}
\begin{array}{l}
m \longmapsto \mu \\
\vm \longmapsto \Am \\
\lamBa \longmapsto \lamBa \\
\lama \longmapsto \lama + \frof \Eab \Cbg 
                  \sigma^{\mu}_{{\gamma}{\dot{\gamma}}}
                  \{ \lamBgu , \Am \} \\
\DC \longmapsto D -i\pME \Amu. \\
\end{array}
\end{equation}
We use Greek indices for Euclidean spacetime and Latin indices for Minkowski spacetime. In \cite{sb03}, only 
products of $\OH$ were deformed. It is clear that we can recover this deformation by setting $\CBab$ to zero wherever it occurs. Using the dictionary and setting $\CBab =0$, we have
\begin{equation}
\begin{array}{l}
e^V = 1 + V + \frot V\ast V \\
 \ \ \ \ \ = 1-\OmOE A_{\mu} +i\OO\OHB\lamB 
 -i\OOB\Oa (\lama +\frof \Eab \Cbg \SmggE \{ \lamBgu , \Am \} ) \\ 
\ \ \ \ \ \ \ \ \ \ \, + \ \frot \OO\OOB \DCA - \frof \CmnE \Am \An \OOB \\
\ \ \ \ \ \ \ \ \ \ \, - \ \frif \Cab \Obd \SmaaE [ \Am , \lamBau ]
          \OOB -\fros |C|^2 \lamB \lamB \OOB. \\  
\end{array}
\end{equation}
This is precisely the exponential that Seiberg found on noncommutative Euclidean superspace in \cite{sb03}.

\section{Gauge theory on $N=0$ Minkowski superspace} \label{S:gauge}

In this section, we generalize super Yang-Mills theory to deformed Minkowski superspace. Most of the usual constructions hold and the approach is similar to Sieberg's $N=\frot$ super Yang Mills theory in \cite{sb03}. We simply replace products in \cite{wb} with star products. The main subtlety is finding the correct parametrization of the vector superfield.


\subsection{Gauge Transformations} \label{S:gauge transformations}

Our goal is to find a way to embed the usual C-independent gauge transformations into superfield equations on noncommutative Minkowski superspace. Since our spinors are built on Minkowski space, we are forced to relate $\OH$ and $\OHB$ by conjugation. This means that we cannot directly follow the construction of \cite{sb03}. In \cite{sb03}, we can see that $\overline{(\Oa)} \ne \OBa $, $\overline{V} \ne V $  and $\overline{( \Lam +\LamB )} \ne \Lam + \LamB $. These relations are sensible for Seiberg, who wrote them over noncommutative Euclidean superspace. On Minkowski space, these inequalities must become equalities. We will find that these reality conditions and the requirement that we recover $N=\frot$ theory in the $\CBab=0$ limit almost uniquely fixes this construction. 


Nonabelian gauge transformations on the vector superfield are embedded into the following superfield equation on noncommutative Minkowski superspace.
\begin{equation} \label{E:expVtransform}
\begin{array}{l}
e^{V} \longmapsto {e^{V}}'=e^{-i\LamB }\ast e^V \ast e^{i\Lam }
\end{array}
\end{equation}
This is the natural modification of \cite{wb}. Infinitesimally, we have
\begin{equation}\label{E:sfadjoint}
\begin{array}{l}
\DT e^V = -i\LamB \ast e^V + ie^V \ast \Lam.
\end{array}
\end{equation}
The component fields of the vector superfield should transform in the adjoint representation of the gauge group as in the standard gauge theory. That is, under an infinitesimal gauge transformation, we should have
\begin{equation}\label{E:adjoint}
\begin{array}{l}
\DT \vm = -2\pM \phi +i[\phi ,\vm ] \\
\DT \lama = i[\phi ,\lama ] \\
\DT D = i[\phi , D]. \\
\end{array}
\end{equation}
Our goal now is to find a suitable parametrization of the gauge parameter $\Lam$ and the vector superfield $V$ such that (\ref{E:adjoint}) are embedded into (\ref{E:sfadjoint}). It is not surprising that the canonical Wess-Zumino gauge (\ref{E:stexp}) does not work in the $N=0$ case, since it was also necessary for \cite{sb03} to shift the $\lam$ component in the $N=\frot$ case. The reality of V requires that we cannot shift only $\lam$; we must also shift $\lamB$. To be precise, $\lam \mapsto \lam + A$ and $\lamB \mapsto \lamB + B$. We now determine what choice of $A$ and $B$ will preserve the reality of $V$ while concurrently embedding (\ref{E:adjoint}). To the first order in $C$, we find under the above redefinitions that (\ref{E:stexp}) becomes,
\begin{equation} \label{E:shiftedstexp}
\begin{array}{l}
e^V = 1-\OmO \vm  -\frof \CBmn \vm \vn \OO
      +\frof \Cmn \vm \vn \OOB +\frot \DC \OO \OOB \\
\ \ \ \ \qquad + \ \OOB \Oa ( -i\lama -iA
      + \frif \Eab \Cbg \Smga [ \vm , \lamBau ] ) \\
\ \ \ \ \qquad + \ \OO \OBa ( -i\lamBa -iB
      -\frif \EBabd \CBbg \Smag [ \vm , \lamau ] ). \\
\end{array}
\end{equation}
Additionally, we make a C-dependent shift of the gauge parameter $\Lam $ similar to that of \cite{sb03}. For the moment, let us make a reasonably general ansatz for the gauge parameter in terms of a variable $p$.
\begin{equation} \label{E:pgauge}
\begin{array}{l}
\Lamp = -\phi +ip\OmO \pMp +\frit \OO \CBmn \{ \vn , \pMp \} 
                           -(p+1)\OO\OOB \ppp \\
\LamBp =-\phi +i(2-p)\OmO \pMp -\frit \OOB \Cmn \{ \pMp , \vn \} 
                           -(p+1)\OO\OOB \ppp \\
\end{array}
\end{equation}
where everything is a function of $y$ in the above. Notice that modulo the higher $\OH$ components in $\Lam$, this reduces to the choice of gauge parameter in \cite{sb03} when $p=0$. We now determine which choice of $p$ will embed (\ref{E:adjoint}) in (\ref{E:sfadjoint}). We calculate that the $\OOB \Oa $ term in the RHS of (\ref{E:sfadjoint}) is
\begin{equation} \label{E:sfadjointL}
\begin{array}{l}
[ \phi,\lama ] + [\phi,A] - \frof\bob \bigl( [\phi, [\lamBau,\vm] ]
               - 2i(p\lamBau \pMp + (2-p) \pMp \lamBau) \bigr). 
\end{array}
\end{equation}
Similarly, the $\OO \OBa $ term in the RHS of (\ref{E:sfadjoint}) is
\begin{equation} \label{E:sfadjointLB}
\begin{array}{l}
[ \phi,\lamBa ] + [\phi,B] + \frof\jane \bigl( [\phi, [\lamau,\vm] ]
                + 2i(p\lamau \pMp + (2-p) \pMp \lamau) \bigr). 
\end{array}
\end{equation}
The $\OOB \Oa $ component of the LHS of (\ref{E:sfadjoint}) is 
\begin{equation} \label{E:varL}
\begin{array}{l}
-i\DT \lama -i\DT A +\frif\bob \DT [ \lamBau,\vm ].
\end{array}
\end{equation}
Similarly, the $\OO \OBa $ component of the 
LHS of (\ref{E:sfadjoint}) is 
\begin{equation} \label{E:varLB}
\begin{array}{l}
-i\DT \lamBa -i\DT B -\frif \jane \DT [ \lamau,\vm ].
\end{array}
\end{equation}
It is not difficult to show (applying (\ref{E:adjoint})) that
\begin{equation} \label{E:VarIdentity}
\begin{array}{c}
i\DT [\lamau, \vm ] + [\phi , [\lamau, \vm ]] = -2i[\lamau, \pMp ] \\
i\DT [\lamBau, \vm ] + [\phi , [\lamBau, \vm ]] = 2i[\lamBau, \pMp ] \\
i\DT \{ \lamBau, \vm \} + [\phi , \{ \lamBau, \vm \} ] = 
                        2i\{\lamBau, \pMp \} \\
i\DT (\vm \lamau ) + [\phi , \vm \lamau ] = 2i\pMp \lamau \\
i\DT (\lamBau \vm ) + [\phi ,\lamBau \vm ] = 2i\lamBau \pMp. \\
\end{array}
\end{equation}
Next, equate (\ref{E:varL}) and (\ref{E:sfadjointL}). Then require that $\DT \lamau = i[ \phi , \lamau ] $ so that (\ref{E:VarIdentity}) holds. Some terms cancel and we find that 
\begin{equation} \label{E:ConA}
\begin{array}{l}
-i\DT A -[\phi , A ] = \frit \bob \bigl( (p+1)\lamBau \pMp
                                    +(1-p)\pMp \lamBau \bigr). 
\end{array}
\end{equation}
Likewise, equate (\ref{E:varLB}) and (\ref{E:sfadjointLB}). Then require that $\DT \lamBau = i[ \phi , \lamBau ] $ so that (\ref{E:VarIdentity}) holds. Some terms cancel and we find that 
\begin{equation} \label{E:ConB}
\begin{array}{l}
-i\DT B -[\phi , B ] = \frit \bob \bigl( (p-1)\lamau \pMp
                                    +(3-p)\pMp \lamau \bigr). 
\end{array}
\end{equation}
When $p=0$, we find that (\ref{E:ConA}) becomes
\begin{equation} 
\begin{array}{l}
-i\DT A -[\phi , A ] = \frit \bob \{ \lamBau , \pMp \}. 
\end{array}
\end{equation}
Hence, in view of (\ref{E:VarIdentity}), we can see why \cite{sb03} shifted the $\lama$ component of the vector multiplet by 
$A=\frof \bob \{ \lamBau , \vm \} $. If we tried to use this choice of gauge parameter, we would destroy the reality of V because (\ref{E:ConB}) would lead us to choose $B= \frof \jane (-\lamau \vm +3\vm \lamau)$. The correct choice is $p=1$. With this choice of gauge parameter, we find the following conditions for $A$ and $B$ from (\ref{E:ConA}) and (\ref{E:ConB}):
\begin{equation} 
\begin{array}{l}
-i\DT A -[\phi , A ] = i\bob \lamBau \pMp \\ 
-i\DT B -[\phi , B ] = i\jane \pMp \lamau. \\ 
\end{array}
\end{equation}
These conditions are satisfied by 
\begin{equation} 
\begin{array}{l}
A=\frot \bob \lamBau \vm \\
B=\frot \jane \vm \lamau. \\
\end{array}
\end{equation}
It is easy to see that $\bar{A} = B$ and $\bar{B} = A$, which is necessary in order to preserve $\bar{V}=V$. This is the only parametrization of the vector superfield and gauge parameter for noncommutative Minkowski superspace if we wish to stay in a generalized Wess Zumino gauge. In principle, we could use the other lower $\OH$ components of the vector superfield to do more complicated shifts. Fortunately, we will not need to do that.  Define the vector superfield to be
\begin{equation} \label{E:vsf}
\begin{array}{l}
V (y) = -\OmO \vm + \OO \OBa ( -i\lamBa +\frit \jane \vm \lamau ) \\
\ \ \ \ \ + \OOB \Oa ( -i\lama -\frit \bob \lamBau \vm )
              +\frot \OO \OOB \DC. 
\end{array}
\end{equation}
It should be evident from the calculations in this section that this parametrization of V embeds (\ref{E:adjoint}) in (\ref{E:sfadjoint})
while maintaining the reality of V. This, of course, requires that we define the gauge parameters as functions of $y$ to be
\begin{equation} \label{E:thegauge}
\begin{array}{l}
\Lam (y) = -\phi +i\OmO \pMp +\frit \OO \CBmn \{ \vn , \pMp \} 
                           -2\OO\OOB \ppp \\
\LamB (y) =-\phi +i\OmO \pMp -\frit \OOB \Cmn \{ \pMp , \vn \} 
                           -2\OO\OOB \ppp. \\
\end{array}
\end{equation}
For the remainder of this paper, we will assume that the vector superfield is parametrized as in (\ref{E:vsf}) and that the gauge parameter is parametrized as in (\ref{E:thegauge}).  Explicitly in this parameterization, 
to the first order in C, (\ref{E:stexp}) becomes: 
\begin{equation} \label{E:stexpP}
\begin{array}{l}
e^V = 1-\OmO \vm  -\frof \CBmn \vm \vn \OO
      +\frof \Cmn \vm \vn \OOB +\frot \DC \OO \OOB \\
\ \ \ \ \qquad + \ \OOB \Oa ( -i\lama 
        - \frif \Eab \Cbg \Smga \{ \lamBau , \vm  \} ) \\
\ \ \ \ \qquad + \ \OO \OBa ( -i\lamBa 
      -\frif \EBabd \CBbg \Smag \{ \lamau , \vm \} ). \\
\end{array}
\end{equation}


\subsection{Spinor superfields} \label{S:field strength}
Again, we will construct these as in the commutative theory except that everywhere that we had a product in the 
commutative theory, we place a star product here. Define
\begin{equation}
\begin{array}{l}
\Wa = -\frof \DBa \ast \DBau \ast e^{-V} \ast \Da \ast e^{V}.   \\
\end{array}
\end{equation}
Conveniently, in chiral coordinates $y^m = x^m + i\OmO$, several of the star products in the above are ordinary products. Thus,
\begin{equation}
\begin{array}{l}
\Wa = -\frof \DBa \DBau e^{-V} \ast \Dau \ast e^{V}. \\
\end{array}
\end{equation}
Likewise define
\begin{equation}
\begin{array}{l}
\WBa = -\frof \Dau \ast \Da \ast e^{-V} \ast \DBa \ast e^{V}.  \\
\end{array}
\end{equation}
Similarly, in antichiral coordinates ${\bar{y}}^m = x^m - i\OmO$, the above simplifies to
\begin{equation}
\begin{array}{l}
\WBa = -\frof \Dau \Da e^{-V} \ast \DBa \ast e^{V}. \\
\end{array}
\end{equation}

We must determine the component field content of $\Wa$ and $\WBa$. Referring to (\ref{E:stexpP}) and keeping only up to the first order in $C$, we obtain
\begin{equation}
\begin{array}{l}
\Wa = \Wa (C=0) \\
\ \ \ \ \ \ \ \ \ + \OO ( \frot \CBmn \{ \Fmn , \lama \}
                  +\CBmn \{ \vn , \covDm \lama 
                  - \frif [ \vm , \lama ] \} ) \\
\ \ \ \ \ \ \ \ \ + \Cgb \Eba \Ogd \lamB \lamB,  \\
\end{array}
\end{equation}
where following Wess and Bagger's conventions in \cite{wb}, the field strength and covariant derivative of the gaugino are
\begin{equation}
\begin{array}{l}
     \Fmn = \pmvn -\pnvm +\frit [ \vm, \vn ] \\
     \covDm \lama = \pM \lama +\frit [ \vm , \lama ].
\end{array}
\end{equation}
Additionally, the spinor superfield of ordinary superspace is
\begin{equation}
\begin{array}{l}
\Wa (C=0) = -i\lama + \Oad D -\Smnab \Obd \Fmn +\OO \Smab \covDm \lamBbu.
\end{array}
\end{equation}
Notice that when we set $\CBab = 0$, we recover the result of Seiberg 
\cite{sb03} for $\Wa$. Likewise, we find that
\begin{equation}
\begin{array}{l}
\WBa = \WBa (C=0) \\
\ \ \ \ \ \ \ \ \ + \OOB ( \frot \Cmn \{ \Fmn , \lamBa \}
                  +\Cmn \{ \vn , \covDm \lamBa 
                  - \frif [ \vm , \lamBa ] \} ) \\
\ \ \ \ \ \ \ \ \ + \CBgb \EBba \OBgd \lambda \lambda  \\
\end{array}
\end{equation}
where
\begin{equation}
\begin{array}{l}
\WBa (C=0) = i\lamBa + \OBad D -\SBmnab \OBbd \Fmn 
                     +\OOB \SBmab \covDm \lambu.
\end{array}
\end{equation}
Again, we reproduce the result of \cite{sb03} upon setting $\CBab = 0$.


\subsubsection{Gauge transformation of spinor superfields} \label{S:Wtransform}
The spinor superfield transforms as in the commutative theory. 
From the nonabelian gauge transformation (\ref{E:expVtransform}), it follows that
\begin{equation} \label{E:Wgauges2}
\begin{array}{l}
\Wa \mapsto {W}_{\alpha}^{'} = e^{-i\LamB }\ast \Wa \ast e^{i\Lam }.
\end{array}
\end{equation}
This can be shown by modifying the calculation used in the commutative theory. We simply change products to star products and utilize the algebra given in (\ref{E:QDalgebra}).


\section{Chiral and antichiral superfields} \label{S:matter fields}
Chiral ($\Phi $) and antichiral($\Phid $) superfields are defined as usual.
\begin{equation}
\begin{array}{ll}
\DBa \ast \Phi = 0  \ \ \ \ \ \ \ &\Da \ast \Phid = 0 \\
\end{array}
\end{equation}
The stars deform any multiplications that result. However, as $\Da = \pa $  in the chiral 
coordinates $y^{\mu} = x^{\mu} +i\OmO $ and $\DBa = \pad $
in the antichiral coordinates ${\bar{y}}^{\mu} = x^{\mu} -i\OmO $, we find that the star products 
are ordinary products. Consequently, we find the well-known solutions
\begin{equation} \label{E:canChiral}
\begin{array}{l}
 \Phi (y, \OH ) = A{(}y{)} + {\sqrt{2} } \OH \psi {(}y{)} + 
                              \OO F{(}y{)} \\
 \Phid (\yb ,\OHB) = 
        \Ab (\yb) + {\sqrt{2} } \OHB \PsiB (\yb ) + \OO \Fb (\yb ).
\end{array}
\end{equation}
These solutions follow from the chain rule as in the standard commutative theory. This construction need not be modified on noncommutative Minkowski superspace because the anticommutation relations given in (\ref{E:QDalgebra}) are uneffected by the deformation.


\subsection{Parametrizing the Chiral superfield} \label{S:chiralpara}
The matter fields in the Wess-Zumino model should transform in the fundamental and antifundamental representations of the gauge group. This is naturally embedded into the following superfield equation written on noncommutative Minkowski superspace, (as T. Araki, K. Ito and A. Ohtsuka did for Euclidean case in \cite{aio}),
\begin{equation} \label{E:bigfun}
\begin{array}{c}
\Phi \mapsto {\Phi}' = e^{-i\Lam}\ast \Phi \ \ \ \ \ \
\PhiB \mapsto {\PhiB}' = \bar{\Phi} \ast e^{i\LamB}. \\
\end{array}
\end{equation}
Infinitesimally, we have
\begin{equation} \label{E:littlefun}
\begin{array}{c}
\DT \Phi =-i\Lam \ast \Phi \ \ \ \ \ \ 
\DT \PhiB = i\PhiB \ast \LamB. \\
\end{array}
\end{equation}
 At the level of component fields, (\ref{E:littlefun}) should embed
\begin{equation} \label{E:compChiral}
\begin{array}{ll}
\DT A(y) = i\phi A(y) & \ \ \ \DT \Ab ( \yb ) = -i \Ab \phi (\yb ) \\
\DT \psi(y) = i\phi \psi(y) & \ \ \ 
    \DT \PsiB ( \yb ) = -i \PsiB \phi (\yb ) \\
\DT F(y) = i\phi F(y) & \ \ \ \DT \Fb ( \yb ) = -i \Fb \phi (\yb ). \\
\end{array}
\end{equation}
It was necessary for \cite{aio} to shift the $\bar{F}$-term in $\PhiB$ to maintain the usual C-independent gauge transformations on the component fields. Similarly, we must modify both $\Phi$ and $\PhiB$ from the cannonical form given in (\ref{E:canChiral}).
\begin{equation} \label{E:modChiral}
\begin{array}{l}
 \Phi (y) = A +\sqrt{2} \OH \psi + \OO (F+\eta) \\
 \Phid (\yb ) = \Ab + \sqrt{2} \OHB \PsiB + \OOB ( \Fb +\beta ) 
\end{array}
\end{equation}
where the shifts $\eta$ and $\beta$ must be chosen as to embed
(\ref{E:compChiral}) in (\ref{E:littlefun}). Now $\Lam$ and $\LamB$ were given in (\ref{E:thegauge}), however, it will be convenient to view $\LamB$ as a function of $\bar{y}$ for this section.
\begin{equation} \label{E:thegauge2}
\begin{array}{l}
\Lam (y) = -\phi +i\OmO \pMp +\frit \OO \CBmn \{ \vn , \pMp \} 
                           -2\OO\OOB \ppp \\
\LamB (\bar{y}) =-\phi -i\OmO \pMp -\frit \OOB \Cmn \{ \pMp , \vn \} 
                           -2\OO\OOB \ppp \\
\end{array}
\end{equation}
The $\OO $ coefficient in (\ref{E:littlefun}) yields
\begin{equation} \label{E:pg157}
\begin{array}{l}
\DT F + \DT \eta = i\phi F +i\phi \eta 
-2i\CBmn \pMp \pN A +\frot \CBmn \{ \vn , \pMp \} A.
\end{array}
\end{equation}
Likewise, the $\OOB$ coefficient in (\ref{E:littlefun}) yields
\begin{equation} \label{E:pg156}
\begin{array}{l}
\DT \Fb + \DT \beta = -i\Fb \phi -i\beta \phi 
-2i\Cmn \pN \Ab \pMp +\frot \Cmn \Ab \{ \pMp ,\vn \}. 
\end{array}
\end{equation}
If we require that (\ref{E:compChiral}) holds, then we then find that the following condition on $\beta$ from (\ref{E:pg156}) is
\begin{equation} \label{E:pg156b}
\begin{array}{l}
\DT \beta -i\phi \beta =-2i\Cmn \pN \Ab \pMp 
                      +\frot \Cmn \Ab \{ \pMp ,\vm \}. 
\end{array}
\end{equation}
Similarly, we find that the following condition on $\eta$ from (\ref{E:pg157}) is
\begin{equation} \label{E:pg158}
\begin{array}{l}
\DT \eta -i\phi \eta =-2i\CBmn \pMp \pN A 
                      +\frot \CBmn \{ \vn ,\pMp \} A.
\end{array}
\end{equation}
Following \cite{aio}, we notice that
\begin{equation} \label{E:pg153}
\begin{array}{l}
\DT [i\Cmn \pM  ( \Ab \vn )   - \frof \Cmn \Ab \vm \vn ]
   +i[i\Cmn ( \pM  \Ab \vn )   - \frof \Cmn \Ab \vm \vn ]\phi = \\
  = -2i\Cmn ( \pM \Ab )(\pNp) + \frot \Cmn \Ab \{ \pMp , \vn \}. \\
\end{array}
\end{equation}
Additionally, we note that
\begin{equation} \label{E:pg159}
\begin{array}{l}
\DT [-i\CBmn \pM \vn A   + \frof \CBmn \vm \vn A ]
   -i\phi[-i\CBmn \pM ( \vn A )   + \frof \CBmn \vm \vn A ] = \\
  = 2i\CBmn (\pNp)( \pM A ) + \frot \CBmn \{ \vn , \pMp \} A. \\
\end{array}
\end{equation}
Then, observe that (\ref{E:pg159}) and  (\ref{E:pg158}) indicate that
\begin{equation} \label{E:Fshift}
\begin{array}{l}
\eta = -i\CBmn \pM ( \vn A )  + \frof \CBmn \vm \vn A. 
\end{array}
\end{equation}
Then, observe that (\ref{E:pg153}) and  (\ref{E:pg156b}) indicate that
\begin{equation} \label{E:FBshift}
\begin{array}{l}
\beta = i\Cmn \pM ( \Ab \vn )  - \frof \Cmn \Ab \vm \vn. 
\end{array}
\end{equation}
Thus, we define the chiral and antichiral superfields 
with respect to (\ref{E:thegauge}) as
\begin{equation} \label{E:shiftedChiral}
\begin{array}{l}
 \Phi  = A +\sqrt{2} \OH \psi + \OO (F
              -i\CBmn \pM (\vn A )   + \frof \CBmn \vm \vn A ) \\
 \Phid  = \Ab + \sqrt{2} \OHB \PsiB + \OOB ( \Fb 
              +i\Cmn \pM (\Ab \vn )  - \frof \Cmn \Ab \vm \vn ). 
\end{array}
\end{equation}
It should be clear from this section that this is the correct 
parametrization of the anti(chiral) superfields. This definition 
embeds (\ref{E:compChiral}) in (\ref{E:littlefun}). This parametrization 
gives the component fields the standard C-independent gauge transformations.


\section{Gauged Wess-Zumino Model} \label{S:lagrangian}
We construct the gauge invariant Lagrangian of the Wess-Zumino model on noncommutative Minkowski superspace:
\begin{equation} \label{E:wzlagrangian}
\begin{array}{l}
  \mathcal{L} = 
     \frac{1}{16kg^2} \biggl( \int d^2 \OH \text{tr} W \ast W 
    +\int d^2 \OHB \text{tr} \WB \ast \WB \biggr)
    +\int  \ d^2 \OH d^2 \OHB \PhiB \ast e^{V} \ast \Phi.
\end{array}
\end{equation}
Gauge invariance of $\mathcal{L}$ follows directly from the cyclicity of the trace and equations (\ref{E:expVtransform}), (\ref{E:Wgauges2}) ,  and (\ref{E:bigfun}). Also, note that this Lagrangian is real as the star product has the property $\overline{f\ast g} = \overline{g} \ast \overline{f}$. To first order in the deformation parameter, we can calculate
\begin{equation} \label{E:puregauge}
\begin{array}{l}
 \text{tr} W \ast W |_{\OO} 
   = \text{tr} W \ast W (C=0) |_{\OO} 
   -i\Cmn \text{tr} \Fmn \lamB \lamB 
   +i\CBmn \text{tr} \lam \lam \Fmn \\
 \text{tr} \WB \ast \WB |_{\OOB} 
   = \text{tr} \WB \ast \WB (C=0) |_{\OOB} 
   -i\Cmn \text{tr} \Fmn \lamB \lamB 
   +i\CBmn \text{tr} \lam \lam \Fmn \\
\end{array}
\end{equation}
where
\begin{equation} \label{E:comm_puregauge}
\begin{array}{l}
  W \ast W (C=0) |_{\OO} 
   = -2i\lamB \SBm \covDm \lambda -\frot \Fmnu \Fmn 
      +D^2 +\frif \Fmnu \Flku \Emnlk \\
  \WB \ast \WB (C=0) |_{\OOB} 
   = -2i\lamB \SBm \covDm \lambda -\frot \Fmnu \Fmn 
      +D^2 -\frif \Fmnu \Flku \Emnlk.
 \end{array}
\end{equation}
To the first order, these terms match those found by \cite{sb03} if we set the $\CBmn=0 $. Next, consider the coupling of the vector and chiral multiplets. After some calculation, we find
\begin{equation} \label{E:gaugekinetic}
\begin{array}{ll}
  \PhiB \ast e^V \ast \Phi|_{\OOBB} 
  =& \Fb F +i\Smaa (\pM \Psibau) \Psiau 
   +\frot \Psibau \Smaa \vm \Psiau  \\
  &+\frot \Ab \DC A - \frof \Ab \vmu \vm A +(\pp \Ab )A \\
  &-i(\pM \Ab ) \vmu A +i\frst \Ab \lambda \psi -i\frst \PsiB \lamB A  \\
  &+i\Cmn \pM (\Ab \vn )F - i\Cmn (\pM \Ab) \vn F  \\
  &-i\CBmn \Fb \pM (\vn A) + i\CBmn \Fb \vn \pM A \\
  &-\frot \Cmn \Ab \vm \vn F +\frot \CBmn \Fb \vm \vn A \\
  &-i\frse \Cab \Smaa \Ab \{ \lamBau , \vm \} \Psib \\
  &-i\frse \CBab \Smaa \PsiBb \{ \lamau , \vm \} A \\
  &-\frst \Cab \Smaa (\pM \Ab ) \lamBau \Psib \\
  &-\frst \CBab \Smaa \PsiBb \lamau \pM A. 
\end{array}
\end{equation}
We identify the terms without deformation parameters as the usual terms in the Wess Zumino model; that is, up to a total derivative we have
\begin{equation} \label{E:comm_gaugekinetic}
\begin{array}{ll}
  \PhiB \ast e^V \ast \Phi (C=0)|_{\OOBB} 
  =& \Fb F -i\PsiB \SBm \covDm \psi -(\covDm \Ab )(\covDmu A) \\
   &+ \frot \Ab DA + \fris (\Ab \lambda \psi - \PsiB \lamB A )
\end{array}
\end{equation}
where $\psi$ and $A$ are in the fundamental representation of the gauge group
\begin{equation} \label{E:covariantfun}
\begin{array}{ll}
  \covDm \psi = \pM \psi + \frit \vm \psi \qquad
  &\covDm A = \pM A + \frit \vm A.
\end{array}
\end{equation}
In (\ref{E:gaugekinetic}), we recover most of the terms found by \cite{aio} plus their conjugates. However, in comparison to the $N=\frot$ theory, terms that are linear in $\lambda$ and $\lamB$ are notably modified. The new shifts
in the gauge parameters (\ref{E:thegauge}) lead to the modification of the 
$\lambda$ and $\lamB$ components of the vector superfield $V$ which in turn give rise to the following terms in the Lagrangian $\mathcal{L}$:
\begin{equation} \label{E:newterms}
\begin{array}{l}
  -i\frse \Cab \Smaa \Ab \{ \lamBau , \vm \} \Psib 
  -\frst \Cab \Smaa (\pM \Ab ) \lamBau \Psib \\
  -i\frse \CBab \Smaa \PsiBb \{ \lamau , \vm \} A 
  -\frst \CBab \Smaa \PsiBb \lamau \pM A. 
\end{array}
\end{equation}
Using covariant derivatives, these terms become
\begin{equation} \label{E:covnewterms}
\begin{array}{l}
  -i\frse \Cab \Smaa \Ab [ \lamBau, \vm ] \Psib
  -\frst \Cab \Smaa (\covDm \Ab ) \lamBau \Psib \\
  +i\frse \CBab \Smaa \PsiBb [ \lamau, \vm ] A
  -\frst \CBab \Smaa \PsiBb \lamau \covDm A. \\
\end{array}
\end{equation}
The term $-\frst \Cab \Smaa (\covDm \Ab ) \lamBau \Psib $ was also found in \cite{aio}. However, the commutator terms result from the choice of gauge parameter we made in (\ref{E:thegauge}). We might naively have expected only the terms without the commutators. Let us summarize:
\begin{equation} \label{E:componentlagrangian}
\begin{array}{ll}
  \mathcal{L} &= \frac{1}{16kg^2} \text{tr} 
     \bigl( -4i\lamB \SBm \covDm \lambda -\Fmnu \Fmn +2D^2 \bigr) \\
     &\ \ \ + \Fb F -i\PsiB \SBm \covDm \psi -\covDm \Ab \covDmu A 
      +\frot \Ab DA + \fris (\Ab \lambda \psi - \PsiB \lamB A ) \\
     &\ \ \ +\frac{1}{16kg^2} \text{tr} 
       \bigl( -2i\Cmn \Fmn \lambda \lambda +2i\CBmn \lamB \lamB \Fmn 
       \bigr) \\
     &\ \ \ +\frit \Cmn \Ab \Fmn F -\frit \CBmn \Fb \Fmn A \\
     &\ \ \ -i\frse \Cab \Smaa \Ab [ \lamBau, \vm ] \Psib
            -\frst \Cab \Smaa (\covDm \Ab ) \lamBau \Psib \\
     &\ \ \ +i\frse \CBab \Smaa \PsiBb [ \lamau, \vm ] A
            -\frst \CBab \Smaa \PsiBb \lamau \covDm A. \\
\end{array}
\end{equation}


\section{Summary}

We have developed a nonabelian gauge theory over deformed Minkowski superspace. In this deformation, all of the fermionic dimensions are deformed and as a result, all of the supersymmetry is broken. To be consistent with the $N=\frot$ terminology, we say that this deformed superspace has $N=0$ supersymmetry. Many of the results directly mirror the results of $N=\frot$ from \cite{sb03} or \cite{aio}. This is due to the fact that the deformation we consider in this paper reduces to the deformation of $N=1/2 $ supersymmetry upon setting $\CBab = 0 $. It is not surprising that we recover the same gauge theoretic results as \cite{sb03} in the limit $\CBab = 0$. The exception to this rule is the choice of gauge parameter introduced by Seiberg in \cite{sb03}. We found that it was not possible to use the same construction because it violated the hermiticity of the vector superfield. We fixed this by introducing a new gauge parameter which served to maintain both hermiticity and the C-independent gauge transformations on the component fields. 

Next, we introduced the chiral superfield $\Phi$. Again, we found it necessary to modify the cannonical component field expansion in order to maintain the standard gauge transformations on the component fields. The modification is similar in spirit to that of \cite{aio}.  Essentially, what we found is the $N=\frot$ theory and conjugate copy where all of the usual $N=\frot$ terms are accompanied by their conjugates due to the hermiticity properties of the star product used in this construction.

Finally, we constructed the Lagrangian which coupled the gauge and matter fields. The gauge invariance of $\mathcal{L}$ follows for reasons similar to the commutative theory. We simply modified the standard arguments for the gauged Wess-Zumino model by replacing products with star products. The primary obstacle to this construction was the task of finding the correct parameterization for the superfields.  The Lagrangian is similar to that found by \cite{aio}, however, there are several new terms. Most new terms come directly from the added deformation $ \{ \OBa , \OBb \}_{\ast} = \CBab $ (which should have been expected from the outset). However, the reparameterization of the gauge parameter also led us to some terms which were not immediately obvious from the $N=\frot$ theory.

There is much work left to do. First, we should complete the program begun in this work to the second order
in the deformation parameter. Nonassociativity will have to be addressed. It is likely that, the constructions 
of this paper will need modification at the second order. Secondly, there are numerous papers investigating
$N=1/2$ supersymmetry \cite{aio}-\cite{billoI} and it would be interesting to find complementary
results for the $N=0$ case where possible. We could try to find the dual results for, instantons as in \cite{ali}-\cite{billo}, or renormalization as in \cite{rey}-\cite{jack} , or the possibility of residual supersymmetry as in \cite{sako}, or the Seiberg Witten map as in \cite{dzo}. We do not attempt to give a complete account of the $N=1/2$ developments, we just wish to point out the variety of novel directions future research might take.   Finally, it would be interesting to derive the $N=0$ deformation from a string theoretical argument.


\section{Appendix} \label{S:appendix}

Define the parity of F to be $\eF$. If F is even, then
$\eF = 1$. If F is odd, then $\eF = -1$. We can express the star product to the first order as:
\begin{equation}
\begin{array}{l}
F \ast G = FG -\frot \Cab \eF ( \Qa F ) ( \Qb G )
              -\frot \CBab \eF ( \QBa F ) ( \QBb G ) \notag.
\end{array}
\end{equation}
Let us then prove that the first order star product is associative. Consider:
\begin{equation}
\begin{array}{ll}
(F \ast G) \ast H 
    &= (FG -\frot \Cab \eF ( \Qa F) ( \Qb G )
           -\frot \CBab \eF ( \QBa F) ( \QBb G )) \ast H \\
    &= FGH -\frot \Cab \eF ( \Qa F ) ( \Qb G )H
           -\frot \CBab \eF ( \QBa F ) ( \QBb G)H \\
           & \qquad \qquad 
           -\frot \Cab \eFG ( \Qa FG ) ( \Qb H)
           -\frot \CBab \eFG ( \QBa FG ) ( \QBb H) \\
    &= FGH -\frot \Cab \eF ( \Qa F ) ( \Qb G )H
           -\frot \CBab \eF ( \QBa F ) ( \QBb G)H \\
           & \qquad \qquad 
           -\frot \Cab \eFG [ (\Qa F)G + \eF F (\Qa G)] \Qb H  \\
           & \qquad \qquad 
           -\frot \CBab \eFG [ (\QBa F)G + \eF  F (\QBa G)] \QBb H  \\
    &= FGH -\frot \Cab \eF ( \Qa F ) ( \Qb G )H
           -\frot \CBab \eF ( \QBa F ) ( \QBb G )H \\
           & \qquad \qquad 
           -\frot \Cab [\eF \eG (\Qa F)G( \Qb H ) 
           + \eG F (\Qa G)( \Qb H )] \\
           & \qquad \qquad 
           -\frot \CBab [\eF \eG (\QBa F)G( \QBb H ) 
           + \eG F(\Qa G)( \QBb H )]. 
\notag
\end{array}
\end{equation}
Notice that we have used $\eFG = \eF \eG $ and $ \eF \eF = 1 $ to complete the calculation above. Likewise consider:
\begin{equation}
\begin{array}{ll}
F \ast (G \ast H) 
  &= F \ast (GH -\frot \Cab \eG ( \Qa G) ( \Qb H )
         -\frot \CBab \eG ( \QBa G) ( \QBb H )) \\
  &= FGH -\frot \Cab \eG F( \Qa G ) ( \Qb H )
         -\frot \CBab \eG F( \QBa G ) ( \QBb H) \\
         & \qquad \qquad 
         -\frot \Cab \eF ( \Qa F ) ( \Qb GH)
         -\frot \CBab \eF ( \QBa F ) ( \QBb GH) \\
  &= FGH -\frot \Cab \eG F( \Qa G ) ( \Qb H )
         -\frot \CBab \eG F( \QBa G ) ( \QBb H) \\
         & \qquad \qquad 
         -\frot \Cab \eF (\Qa F)[(\Qb G)H +\eG G (\Qb H) ] \\
         & \qquad \qquad 
         -\frot \CBab \eF (\QBa F)[(\QBb G)H +\eG G (\QBb H) ] \\
  &= FGH -\frot \Cab \eG F( \Qa G ) ( \Qb H )
         -\frot \CBab \eG F( \QBa G ) ( \QBb H) \\
         & \qquad \qquad 
         -\frot \Cab [\eF (\Qa F)(\Qb G)H +\eF \eG (\Qa F) G (\Qb H)] \\
         & \qquad \qquad 
         -\frot \CBab [\eF (\QBa F)(\QBb G)H +\eF \eG (\Qa F) G (\QBb H) ]. 
         \notag \\
\end{array}
\end{equation}
Therefore, $F \ast (G \ast H) = (F \ast G) \ast H $ to the first order in the deformation parameter.


\section{Acknowledgments}
I would like to thank Daniel Robles-Llana, Manuela Kulaxizi and Ronald Fulp for helpful discussions. Finally, I thank my wife for her considerable patience.

\end{document}